\documentstyle[12pt]{article}
\def\mathbf{\vec}

\def\ca{\c{c}\~{a}}

\begin{document}

\centerline {\LARGE One-Loop Determinant of Dirac Operator in}
\vspace{.3cm}

\centerline {\LARGE Non-Renormalizable Models}
\vspace{1cm}

\centerline {\large A.A. Osipov, B. Hiller, A.H. Blin}
\vspace{.5cm}

\centerline {\it Centro de F\'{\i}sica Te\'{o}rica, Departamento de
F\'{\i}sica}

\centerline {\it da Universidade de Coimbra, 3004-516 Coimbra, Portugal}
\vspace{1cm}

\begin{abstract}
We use proper-time regularizations to define the one-loop fermion determinant 
in the form suggested by Gasser and Leutwyler some years ago. We show how to 
obtain the polynomial by which this definition of $\ln\det D$ needs to be 
modified in order to arrive at the fermion determinant whose modulus is 
invariant under chiral transformations. As an example it is shown how the 
fundamental symmetries associated with the NJL model are preserved in a
consistent way.
\end{abstract}

%%%%%%%%%%%%%%%%%%% PART 1 %%%%%%%%%%%%%%%%%%%%%%%%%%%%%%%%%%%%%%%%%%%%%%%%%%

\section{INTRODUCTION}

In the path-integral formulation of quantum theory the effective action
involves, after integration over the fermionic fields, 
the functional determinant of differential Dirac operator $D$ in the
presence of external sources. The central object in the
calculation of the effective action is always the quantity $\ln\det D$.
We start with the definition
\begin{equation}
\label{logdet}
\ln\det D=-\frac{1}{2}\int^\infty_0\frac{dT}{T}\rho (T,\Lambda^2)\mbox{Tr}
          \left(e^{-T\bar D^2}\right)-\int d^4x P(v,a,\sigma ,\pi ).
\end{equation}
which allows the Schwinger proper-time method to be applied to fermions,
involving the square of the operator $\bar D$. 
Here $\bar D\equiv\gamma_5 D$ has been introduced in 
\cite{Gasser:1984}, $D$ is the Dirac operator in the presence of external 
vector $(v_\mu )$, axial-vector $(a_\mu )$, scalar $(\sigma )$ and pseudoscalar
$(\pi )$ sources. 
This definition of $\ln\det D$ allows to treat the real and imaginary parts of 
$\ln\det D$ on equal footing, as opposite to the $D^{\dagger}D$ definition. 
The polynomial $P(v,a,\sigma,\pi)$, which depends only on the external fields,
is fixed by requiring the modulus of the fermion determinant to be invariant 
under chiral transformations. It has been worked out in the context of a
renormalizable theory \cite{Gasser:1984}. The present work represents an 
extension of the results of \cite{Gasser:1984} to the case
of non-renormalizable models and in particular to incorporate explicitly
the process of dynamical chiral symmetry breaking of the Nambu -- Jona-Lasinio
(NJL) model \cite{Nambu:1961}. 
As an alternative method one can use the 
integral representation of the complex power for the pseudo differential 
operator \cite{Salcedo:1996}. In the latter case an unambiguous definition of 
the determinant of the Dirac operator is obtained. The determinant is shown to 
be vector gauge invariant and to yield the correct axial and scale anomalies.

We consider a class of regularization schemes (proper-time regularizations) 
which can be incorporated in this expression through the kernel $\rho (T, 
\Lambda^2)$. These regularizations allow to shift in loop momenta. A typical 
example is the proper-time cutoff where the kernel $^t\rho (T,\Lambda^2)$ is 
equal to
\begin{equation}
\label{pt}
^t\rho (T,\Lambda^2)=\Theta\left(T-\frac{1}{\Lambda^2}\right).
\end{equation}
Another choice for the kernel can be the covariant Pauli-Villars cutoff
\cite{Pauli:1949}
\begin{equation}
\label{cc}
      ^c\rho (T, \Lambda^2)=1-(1+T\Lambda^2)e^{-T\Lambda^2}
\end{equation}
which leads to the well-known effective potential of the 
NJL model \cite{Nambu:1961}. The result is\footnote{See, for instance,
\cite{Ying:1996} and references in it.}
\begin{equation}
\label{ep}
V(m)=\frac{m^2}{2G}\left(1-\frac{N_cG\Lambda^2}{4\pi^2}\right)+
     \frac{N_c}{8\pi^2}\left[m^4\ln\left(1+\frac{\Lambda^2}{m^2}\right)-
     \Lambda^4\ln\left(1+\frac{m^2}{\Lambda^2}\right)\right].
\end{equation}
Both of the kernels (\ref{pt}) and (\ref{cc}) have been used in many papers,
for example, see papers \cite{Bijnens:1994,Nikolov:1996} and
\cite{Bernard:1992} correspondingly. A wide set of possibilities for the 
kernel $\rho (T, \Lambda^2)$ have been considered in the papers
\cite{Ball:1989,Doring:1992}.

The counterterms $P(v,a,\sigma ,\pi )$ in formula (\ref{logdet}) can be fixed 
from the transformation properties of $\ln\det D$. We consider here the case 
of chiral gauge theories with the $SU(2)_L\times SU(2)_R\times U(1)_V$
chiral symmetry. Explicitly, let $D$ be equal to
\begin{equation}
\label{DD}
D =\gamma^{\mu}(i\partial_{\mu}+v_{\mu}+a_{\mu}\gamma_5) - \sigma
   + i\gamma_5\pi
\end{equation}
where $v_{\mu}=v_{\mu}^i\tau_i,\ a_{\mu}=a_{\mu}^i\tau_i,\ \pi =\pi^a\tau_a,
\ \sigma =\sigma^a\tau_a,\ \tau_a=(1,\tau_i ),\quad [\tau_i, \tau_j]=2i
\epsilon_{ijk}\tau_k, \ i=1,2,3.$ The corresponding chiral transformations of 
the external fields are given by
\begin{equation}
\delta v_\mu = \partial_\mu\alpha +i[\alpha , v_\mu ]+i[\beta , a_\mu ]
\end{equation}
\begin{equation}
\delta a_\mu = \partial_\mu\beta +i[\alpha , a_\mu ]+i[\beta , v_\mu ]
\end{equation}
\begin{equation}
\label{delsig}
\delta\sigma = i[\alpha , \sigma ]-\{\beta , \pi\}
\end{equation}
\begin{equation}
\label{delpi}
\delta\pi = i[\alpha , \pi ]+\{\beta , \sigma\}.
\end{equation}
Here $\alpha = \alpha_i\tau_i$ is the infinitesimal transformation generated by
the vector currents and $\beta = \beta_i\tau_i$ is a chiral transformation.
The transformation law of $\ln\det D$ in this case is known explicitly
\cite{Bardeen:1969} :
\begin{equation}
\label{anomaly}
\delta\ln\det D=\frac{iN_c}{(4\pi )^2}\int
d^4x\mbox{Tr}_f(\beta\Omega )
\end{equation}
where
\begin{eqnarray}
\Omega
&=&\varepsilon^{\alpha\beta\mu\nu}\left[v_{\alpha\beta}v_{\mu\nu} +
          \frac{4}{3}\nabla_{\alpha}a_\beta\nabla_{\mu}a_\nu +
          \frac{2i}{3}\{v_{\alpha\beta}, a_\mu a_\nu\}\right.\nonumber \\
&+&\left. \frac{8i}{3}a_\mu v_{\alpha\beta}a_\nu +\frac{4}{3}a_\alpha a_\beta
          a_\mu a_\nu\right].
\end{eqnarray}
The field strength tensor $v_{\mu\nu}$ associated with $v_\mu$ is defined as
\begin{equation}
v_{\mu\nu}=\partial_\mu v_\nu -\partial_\nu v_\mu -i[v_\mu , v_\nu ]
\end{equation}
and $\nabla_\mu a_\nu$ stands for
\begin{equation}
\nabla_{\mu}a_\nu =\partial_{\mu}a_\nu -i[v_{\mu}, a_\nu ].
\end{equation}

Our aim now is to calculate the polynomial $P(v,a,\sigma ,\pi )$ in the
framework of a nonrenormalizable aproach. Let us note that $P(v,a,\sigma ,\pi 
)$ is unique up to a chirally invariant polynomial. One can always chose $P$ 
in such a manner that the determinant is not modified if the external fields
$a_\mu$ and $\pi$ are switched off. In paper \cite{Gasser:1984} it has been 
shown how to do this for renormalizable theories. There are two essential 
differences in our case. The first one is that we have to use a
regularization with finite cutoff $\Lambda$. The $\zeta$-function technique 
used in \cite{Gasser:1984} is
not good for that because it does not lead to the correct description of the
spontaneous chiral symmetry breaking phenomena. The second one is also related
to the cutoff dependense of the result. As we shall show, the polynomial 
$P(v,a,\sigma ,\pi )$ gets now systematically contributions from the terms 
which would vanish in the limit $\Lambda\rightarrow\infty$. This fact renders 
its evaluation rather technical.

%%%%%%%%%%%%%%%%%%% PART 2 %%%%%%%%%%%%%%%%%%%%%%%%%%%%%%%%%%%%%%%%%%%%%%%%%%

\section{COUNTERTERMS AND SYMMETRY}

To illustrate our consideration we shall discuss the NJL model with the
$SU(2)_L\times SU(2)_R$ chiral symmetry. We use the model version with only
the scalar-scalar and pseudoscalar-pseudoscalar type of four quark 
interactions. Integrating out the quark fields one obtains the action of the 
model in terms of scalar $\sigma\times 1$ and pseudoscalar $\pi =\pi_i\tau^i$ 
collective mesonic degrees of freedom
\begin{equation}
\label{action}
S_{coll} = -i\ln\det\, D-\int d^4 x{(\sigma +m)^2 +\vec\pi^2\over 2G}.
\end{equation}
The Dirac operator $D$ is given by
\begin{equation}
\label{D}
D = i\gamma^{\mu}\partial_{\mu} - m - \sigma + i\gamma_5\pi ,
\end{equation}
where {\it m} denotes the constituent quark mass generated in the process of
spontaneous chiral symmetry breaking. Now in order to be able to derive the
polynomial $P(\sigma,\pi)$, it is crucial to perform the symmetry transformations
in the broken phase. In the phase with broken chiral symmetry
the transformations (\ref{delsig}) and (\ref{delpi}) become
\begin{equation}
\label{delsig2}
\bar{\delta}\sigma = -\{\beta , \pi\}
\end{equation}
\begin{equation}
\label{delpi2}
\bar{\delta}\pi = i[\alpha , \pi ]+\{\beta , \sigma +m\}
\end{equation}
for the considered isospin content of scalar and pseudoscalar. 
Note that if one would first derive $P(\sigma,\pi)$ in the symmetric phase
and then perform the shift $\sigma\rightarrow(\sigma + m)$, it would be
necessary to calculate all orders of the proper-time expansion. All of them 
would contribute as a factor with a certain power $m$ to a fixed order in
the fields. Therefore by constructing the symmetry transformations in the
broken phase (\ref{delsig2}), (\ref{delpi2}), one achieves a resummation of
an infinite number of terms of the symmetric phase.

Under global
chiral transformations the change in the Dirac operator
$\bar{D}=\gamma_5D$ is given by
\begin{equation}
\label{transD}
i\bar{\delta}\bar{D}=[\bar{D}, \alpha ]+\{\bar{D}, \beta\gamma_5\}.
\end{equation}
Therefore, to get the related polynomial $P(\sigma , \pi )$ for this case one 
has to integrate the equality
\begin{equation}
\bar{\delta}\ln\det D=0
\end{equation}
where $\ln\det D$ is defined according to Eq.(\ref{logdet}). The variation of
$P(\sigma , \pi )$ has to cancel the symmetry breaking part coming from
the proper-time integral. In this way one gets
\begin{equation}
\label{varP}
\bar{\delta}P(\sigma , \pi )=\frac{-i}{8\pi^2}\sum^\infty_{n=0}R_n
                             \mbox{tr}(\beta\gamma_5a_{n+1})
\end{equation}
where tr represents trace in internal space. In the case under consideration it
includes summations over flavour, colour and Lorentz indexes:
$\mbox{tr}=\mbox{tr}_f\mbox{tr}_c\mbox{tr}_L$. One can see that $P(\sigma ,
\pi )$ is not invariant under chiral transformations, picking up the 
contribution which is linear in $\beta$. The functions $R_n$ represent the 
integrals which appear in the result of the asymptotic expansion of the heat 
kernel
\begin{eqnarray}
\label{rn}
R_n&=&-\int^\infty_0\rho(T,\Lambda^2)d\left[T^{n-1}e^{-Tm^2}\right]\nonumber \\
   &=&\int^\infty_0 dTT^{n-2}[m^2T-(n-1)]e^{-Tm^2}\rho(T,\Lambda^2).
\end{eqnarray}
These integrals yield the following expression for $R_n$
\begin{equation}
\label{rncc}
^{c}R_n=\frac{n!\Lambda^4}{(\Lambda^2+m^2)^{n+1}}.
\end{equation}
This result corresponds to the kernel (\ref{cc}). For the case of $\rho
(T,\Lambda^2)$ being equal to (\ref{pt}) one gets
\begin{equation}
\label{rnpt}
^{t}R_n=(\Lambda^2)^{1-n}\exp\left(-{m^2\over \Lambda^2}\right).
\end{equation}
In renormalizable theory the terms $R_n$ with $n\geq 2$ vanish in the limit
$\Lambda\rightarrow\infty$. The same is also true if one applies the $\zeta$
- function regularization. This property of renormalizable models extremely 
simplifies the problem. In non-renormalizable models all of $R_n$ terms 
contribute to the result.

The coefficients $a_n\equiv a_n(x,x)$ are the coincidence limit of the Seeley
-- DeWitt coefficients \cite{Ball:1989}. For our illustration we shall need
the first four of them
\begin{eqnarray}
\label{coeff}
a_0&=&1,  \nonumber \\
a_1&=&-Q, \nonumber \\
a_2&=&\frac{1}{2}Q^2+\frac{1}{6}Q_{\mu\mu}+\frac{1}{12}F^2,\nonumber \\
a_3&=&-\frac{1}{6}Q^3-\frac{1}{12}(\{Q,Q_{\mu\mu}\}+Q_\mu^2)-\frac{1}{60}
      Q_{\mu\mu\nu\nu}+\frac{1}{60}[F_{\mu\alpha ;\alpha},Q_\mu]\nonumber \\
   &-&\frac{1}{60}(2\{F^2,Q\}+F_{\mu\nu}QF_{\mu\nu})-\frac{1}{45}F_{\mu\alpha ;
      \alpha}F_{\mu\beta ;\beta}-\frac{1}{180}F_{\mu\alpha ;\beta}F_{\mu\alpha
      ;\beta }\nonumber \\
   &-&\frac{1}{60}\{F_{\mu\nu}, F_{\mu\nu ;\alpha\alpha}\}+\frac{1}{30}F^3.
\end{eqnarray}
Some comments are in order here. First, we deal in this case with the linear
realization of chiral symmetry. It means that we have for $\bar D^2$ the
following representation
\begin{equation}
\bar D^2=\nabla_{\mu}\nabla^{\mu}+m^2+Q
\end{equation}
where
\begin{equation}
\nabla_{\mu}=\partial_{\mu}+A_{\mu}, \quad
             A_{\mu}=\gamma_{\mu}\gamma_5\pi ,
\end{equation}
\begin{equation}
Q=(\sigma^2+2m\sigma )+i\gamma^{\mu}\partial_{\mu}\sigma -2
             (m+\sigma )i\gamma_5\pi +3\vec\pi^2.
\end{equation}
Second, we wrote the coefficients (\ref{coeff}) directly in Minkowski
space. In this way one should understand summations over repeated Lorentz
indexes to be implicit. We have used the following designations
\begin{equation}
F_{\mu\nu} = [\nabla_{\mu}, \nabla_{\nu}]=\gamma_{\nu}\gamma_5
             \partial_{\mu}\pi-\gamma_{\mu}\gamma_5\partial_{\nu}\pi
             +[\gamma_{\nu}, \gamma_{\mu}]\vec\pi^2,
\end{equation}
\begin{equation}
F^2 = F_{\mu\nu}F^{\mu\nu}, \quad
F^3=F_{\mu\nu}F^{\nu\sigma}F_{\sigma\mu},
\end{equation}
\begin{equation}
Q_{\mu}=[\nabla_{\mu}, Q], \quad F_{\mu\nu;\nu}=[\nabla_{\nu},
F_{\mu\nu}],
\end{equation}

One has to calculate traces $\mbox{tr}(\beta\gamma_5 a_n)$ and integrate
Eq.(\ref{varP}). The first three non-zero contributions are given by
\begin{equation}
\label{a1}
\mbox{tr}(\beta\gamma_5 a_1)=4iN_c\bar{\delta}\vec\pi^2.
\end{equation}
\begin{equation}
\mbox{tr}(\beta\gamma_5 a_2)=4iN_c\bar{\delta}\left[\frac{1}{6}(\partial_\mu
\vec\pi )^2-2m\sigma\vec\pi^2-\sigma^2\vec\pi^2-\frac{2}{3}\vec\pi^4\right].
\end{equation}
\begin{eqnarray}
\mbox{tr}(\beta\gamma_5 a_3)&=&4iN_c\bar{\delta}\left[\frac{1}{60}
(\partial_\mu^2\vec\pi )^2-\frac{1}{2}\vec\pi^2(\partial_\mu\sigma)^2
-\frac{1}{5}\vec\pi^2(\partial_\mu\vec\pi )^2-\frac{7}{60}
(\partial_\mu\vec\pi^2 )^2 \right.\nonumber \\
   &-&\frac{1}{3}(\sigma +m)\partial_\mu\sigma\partial_\mu\vec\pi^2
      +\frac{1}{10}\vec\pi^6 \nonumber \\
   &+&\frac{1}{6}(\sigma^2+2m\sigma )[3\vec\pi^2 (\sigma^2+2m\sigma
      )+2\vec\pi^4 -(\partial_\mu\vec\pi )^2] \nonumber \\
   &-&\left.\frac{1}{3}m^2\vec\pi^4-2m\sigma\vec\pi^2-\sigma^2\vec\pi^2
      -\frac{2}{3}\vec\pi^4\right].
\end{eqnarray}
Let us note that the last four terms from $a_3$ (see Eq.(\ref{coeff})) do not
contribute to $\mbox{tr}(\beta\gamma_5 a_3)$.

On the other side, the first term in formula (\ref{logdet}) contributes to the
Lagrangian of collective fields as
\begin{equation}
\label{lcoll1}
{\cal L}_{coll}^{(1)}=-\frac{1}{32\pi^2}\sum^\infty_{n=0}J_n\mbox{tr}
                      (a_{n+1}),
\end{equation}
where
\begin{equation}
J_n=\int^\infty_0\frac{dT}{T^{2-n}}e^{-Tm^2}\rho (T,\Lambda^2),
         \quad n=0,1,2...
\end{equation}
We have from (\ref{lcoll1})
\begin{eqnarray}
{\cal L}_{coll}^{(1)}&=&\frac{N_c}{(2\pi )^2}\left\{(\sigma^2+2m\sigma
              +3\vec\pi^2)J_0\right. \nonumber \\
     &+&\left.\frac{1}{2}\left[(\partial_\mu\sigma )^2+(\partial_\mu\vec\pi
      )^2+4m^2\vec\pi^2-(\sigma^2+2m\sigma +\vec\pi^2)^2\right]J_1+...\right\}.
\end{eqnarray}
Using the identity
\begin{equation}
m^2J_n+(2-n)J_{n-1}=R_{n-1}\quad n=1,2,3...
\end{equation}
one can see, for instance, how symmetry breaking terms proportional to
$\vec\pi^2$ are compensated in this expression by the contribution from 
(\ref{a1}). A fully chiral symmetric Lagrangian is therefore obtained at 
each order of the proper time expansion.

%%%%%%%%%%%%%%%%%%% PART 3 %%%%%%%%%%%%%%%%%%%%%%%%%%%%%%%%%%%%%%%%%%%%%%%%%%

\section{CONCLUSIONS}

We used the one-loop fermion determinant in the form suggested by Gasser and
Leutwyler some years ago \cite{Gasser:1984} to extend it to be applicable to
non-renormalizable models. In this way the real and imaginary parts of
$\ln\det D$ can be calculated with the same input. One obtains the correct 
description of the chiral anomaly when regularization is switched off. However 
it is necessary to correct the real part of $\ln\det D$ by the polynomial
$P(v,a,\sigma ,\pi )$ to get the chiral invariant result for this case. We 
have shown how to get the chiral symmetry restoring polynomial $P(v,a,\sigma 
,\pi )$. The simplest way to do this is to calculate in the phase with broken 
chiral symmetry, rewriting the symmetry transformations especially for this 
case. The result is an extension of the form presented in the paper 
\cite{Gasser:1984}, to incorporate explicitly the process of dynamical chiral 
symmetry breaking of the NJL model.

For simplicity we have considered the NJL model without vector and axial-vector
degrees of freedom. However, the result can be easily extended to the more
general case.

%%%%%%%%%%%%%%%%%%%%%%%% Acknowledgement %%%%%%%%%%%%%%%%%%%%%%%%%%%%%%%%%%%%%%

\section{ACKNOWLEDGEMENTS}

This work is supported by grants provided by Funda\ca o para a Ci\^encia e a
Tecnologia, PRAXIS XXI/BCC/7301/96, PRAXIS/C/FIS/12247/1998,
PESO/P/PRO/15127/1999 and NATO "Outreach" Cooperation Program.

%%%%%%%%%%%%%%%%%% REFERENCES %%%%%%%%%%%%%%%%%%%%%%%%%%%%%%%%%%%%%%%%%%%%%%%%%
\baselineskip 12pt plus 2pt minus 2pt

\end{document}